\documentclass[final,authoryear,5p,times,twocolumn]{elsarticle}
\pdfoutput=1

\usepackage{graphicx}
\usepackage{amssymb}
\usepackage[,pdftex,pdfpagemode={UseOutlines},bookmarks,bookmarksopen,colorlinks,linkcolor={blue},citecolor={green},urlcolor={red}]{hyperref}
\usepackage{hypernat}

\journal{Astronomy \& Computing}

\usepackage{upquote}

\usepackage{upgreek}

\usepackage{color}

\usepackage{listings}

\definecolor{mygreen}{rgb}{0,0.6,0}
\definecolor{mygray}{rgb}{0.5,0.5,0.5}
\definecolor{mymauve}{rgb}{0.58,0,0.82}

\newcommand{\ascl}[1]{\href{http://www.ascl.net/#1}{ascl:#1}}

\begin{document}

\begin{frontmatter}



\title{The General Single-Dish Data Format: A Retrospective\tnoteref{ascl}}
\tnotetext[ascl]{These codes are registered at the
Astrophysics Source Code Library (ASCL) with the code entries
\ascl{1503.009} and \ascl{1503.007}.}


\author[jac]{Tim Jenness\corref{cor1}\fnref{timj}}
\ead{tjenness@lsst.org}
\author[noao]{Elizabeth~B.~Stobie}
\author[nrao]{Ronald~J.~Maddalena}
\author[nraocv]{Robert~W.~Garwood}
\author[hp]{Jon~H.~Fairclough}
\author[nrao]{Richard~M.~Prestage}
\author[leiden,imapp]{Remo~P.~J.~Tilanus}
\author[mrao]{Rachael~Padman}

\cortext[cor1]{Corresponding author}
\fntext[timj]{Present address: LSST Project Office, 933 N.\ Cherry Ave, Tucson, AZ 85721, USA}

\address[jac]{Joint Astronomy Centre, 660 N.\ A`oh\=ok\=u Place, Hilo, HI
  96720, USA}
\address[noao]{National Optical Astronomy Observatory, 950 N Cherry Ave, Tucson, AZ 85719, USA}
\address[nrao]{National Radio Astronomy Observatory, P.O.\ Box 2, Green Bank, WV~24944, USA}
\address[nraocv]{National Radio Astronomy Observatory,
  Charlottesville, VA~22903-2475, USA}
\address[hp]{Hewlett Packard Ltd}
\address[leiden]{Leiden Observatory, Leiden University, PO Box 9513, 2300 RA Leiden, The~Netherlands}
\address[imapp]{Department of Astrophysics,
     Institute for Mathematics, Astrophysics and Particle Physics,
     Radboud University Nijmegen, PO Box 9010, 6500 GL Nijmegen, The~Netherlands}
\address[mrao]{Mullard Radio Astronomy Observatory,
Cavendish Laboratory, University of Cambridge,
JJ~Thomson~Avenue,
Cambridge, CB3~0HE, UK}

\begin{abstract}

The General Single-Dish Data format (GSDD) was developed in the
mid-1980s as a data model to support centimeter, millimeter and submillimeter
instrumentation at NRAO, JCMT, the University of Arizona and IRAM. We
provide an overview of the GSDD
requirements and associated data model, discuss the implementation
of the resultant file formats, describe its usage in the observatories and
provide a retrospective on the format.

\end{abstract}


\end{frontmatter}

% \linenumbers

%% Journal abbreviations
\newcommand{\mnras}{MNRAS}
\newcommand{\aap}{A\&A}
\newcommand{\aaps}{A\&AS}
\newcommand{\pasp}{PASP}
\newcommand{\apj}{ApJ}
\newcommand{\apjs}{ApJS}
\newcommand{\qjras}{QJRAS}
\newcommand{\an}{Astron.\ Nach.}
\newcommand{\ijimw}{Int.\ J.\ Infrared \& Millimeter Waves}
\newcommand{\procspie}{Proc.\ SPIE}
\newcommand{\aspconf}{ASP Conf. Ser.}

%% Applications

%% Misc

%% Links


\section{Introduction}

In the late 1970s and early 1980s millimeter and submillimeter
single-dish astronomy was undergoing a significant period of growth
\citep[see e.g.,][]{2013ASSP...37...39R} with the National Radio
Astronomy Observatory (NRAO) 12-m telescope
leading the way \citep[see e.g.,][]{2005ASSL..323.....G} and with
multiple observatories being developed such as the Institut de
Radioastronomie Millim\'{e}trique (IRAM) 30-m
\citep{1981MitAG..54...61B}, the 15-m James Clerk Maxwell Telescope
\citep[JCMT;][]{1985ESOC...22...63H}, the 10-m Sub-Millimeter
Telescope \citep[SMT;][]{1985ESOC...22...71W}, the 15-m Swedish
European Southern Obsevatory Submm Telescope \citep[SEST;][]{1985ESOC...22...25D}, and the Caltech
Submillimeter Observatory \citep[CSO;][]{1988BAAS...20..690P}. In this
environment it was recognized by some institutions that the ability
for raw, or partially processed data taken on one telescope, to be
reduced and analyzed by the software written at another telescope
would be extremely useful and could lead to significant savings on
software development effort.

At this time the Flexible Image Transport System
\citep[FITS;][]{1981A&AS...44..363W} was considered mainly
suitable as a means of exchanging image data using tapes
\citep{1980SPIE..264..298G}. The FITS standard, which then
lacked the capability to use binary tables and could only store a
single ASCII table per file, was not deemed an
efficient format to store complex mm/submm time-series and
spectral-line data from single-dish telescopes that usually required
many sets of tabular data.

The General Single Dish Data format (GSDD) was developed in the 1980s
to solve the data processing and acquisition requirements of the NRAO,
IRAM, University of Arizona and JCMT observatories.  Initial
discussions between NRAO 12m and IRAM staff began in 1983, and
subsequently included JCMT representatives. At around this same time,
however, IRAM started development of the Continuum and Line Analysis
Single-dish Software
\htmladdnormallinkfoot{\textsc{class}}{http://www.iram.fr/IRAMFR/GILDAS}
\citep[][\ascl{1305.010}]{2005sf2a.conf..721P} data reduction package,
and they did not follow up on the GSDD initiative.\footnote{The
  authors have been unable to find anyone from IRAM or the University
  of Arizona that recalls the GSDD discussions and can provide
  information from their side. JCMT and NRAO documents confirm the
  additional parties but no meeting minutes are available. An IRAM
  memo from January 1983 indicates they were strongly in favor of a
  FITS variant called IRAM Disk-oriented FITS (IDFITS) that supported
  VAX floating point, array header keywords, variable length headers
  (up to 80 characters) and \texttt{CONTINUE} cards and stored the
  data in separate files from the header information.}  The GSDD
format, agreed in 1986 \citep[see e.g.,][]{mtdn84,1987NRAO30},
consisted of a data model for specifying centimeter, millimeter and
submillimeter observations (continuum and spectral-line
instrumentation) and a specification of how the bytes would be
represented on disk.  The format was described in both JCMT technical
notes \citep{mtdn84,mtdn85,SUN229} and an NRAO Newsletter article
\citep{1987NRAO30}, but a formal definition of the format was not
published in the literature. In this article we present the first
joint NRAO/JCMT description of the model and provide a retrospective
on the history and usage of the format.
A basic introduction to millimeter and submillimeter observing
techniques is beyond the scope of this paper but good background
information can be provided by \citet{2002ASPC..278.....S}.

\section{Data Model}
\label{sec:datamodel}

To allow interoperability of data files between differing
observatories it was important to develop a shared data model. The
initial approach was to define the simplest possible model to allow
sharing of
raw, or partially reduced spectra between multiple data reduction software
packages. Since JCMT was still in the development phase during these
discussions the focus became how to represent the raw instrument data
on disk. This was simplified somewhat by the JCMT system not storing
individual on-source and off-source or calibration data, but storing
calibrated spectra from the heterodyne systems and chop subtracted
time-series for the continuum instruments.

The model was designed to handle general sub-mm observing techniques
using different switching techniques, such as position switching, beam
switching and frequency switching, and included on-the-fly mapping
techniques (where the telescope is moved during acquisition) as well
as stare and gridded observations.

\begin{table}
\caption{Base classes defined for GSDD. The final two classes listed,
  14 and 55, were only defined at JCMT.}
\label{tab:classes}
\begin{center}
\begin{tabular}{ll}
\hline
Class number &Class name \\ \hline
1         & Basic Information \\
2         & Pointing Parameters  \\
3         & Observing Parameters \\
4         & Positions \\
5         & Environment \\
6         & Map Parameters \\
7         & Data Parameters \\
8         & Engineering Parameters \\
9         & Telescope Dependent Parameters \\
10        & Open Data Reduction Parameters \\
11        & Phase Block \\
12        & Receiver configurations\\
13        & Data Values \\
14        & Pointing History (JCMT) \\
55        & Inclinometry (JCMT)\\
\hline
\end{tabular}
\end{center}
\end{table}

When designing the model related items were grouped into numbered
classes and the parameter name was prefixed by that class number. The
class groupings are shown in Table~\ref{tab:classes}. In early JCMT
documents \citep[e.g.][]{mtdn1,mtdn84} there is disagreement in the class
numbering, for example using \texttt{S2EPH}\footnote{In early
  iterations \texttt{S} was used to indicate a scalar item and
  \texttt{V} a vector item, followed by the class number.} or \texttt{C3EPH} for the epoch of the
coordinates rather than \texttt{C4EPH}, reflecting the uncertainty in
the standardized model, but eventually \citep[see e.g.][]{mtdn85}
the NRAO convention was adopted and the core model solidified
\citep[][defined the NRAO naming scheme]{tcus5}.
Seventy one data items were defined in the shared NRAO/JCMT GSDD data
model.\footnote{72 if the telescope-specific \texttt{C9OT}, Observing
  Tolerance, item is included which was present in the NRAO 12m
  definition and at JCMT but not used for Green Bank. In some very
  early files JCMT erroneously used \texttt{C90T} for this item due to
  a transcription error confusing the letter ``O'' with the number
  zero. This sometimes implies that JCMT used class 90.} and those are
detailed in Table~\ref{tab:core}. For example \texttt{C3DAT} referred
to the UT date of the observation, \texttt{C1SNO} the scan number,
and \texttt{C7VR} the source radial velocity.

\begin{table*}[!ht]
\centering
\caption{Core components of GSDD data model present in both NRAO and
  JCMT implementations \citep{tcus5}. Relevant units are given in square brackets
  using the NRAO convention. JCMT data files contain unit information explicitly.}
\label{tab:core}
\begin{tabular}{|lp{1.5in}|lp{1.5in}|lp{1.5in}|}
\hline
\textbf{C1BKE} & Backend & \textbf{C4CSC} & Code for coordinate system & \textbf{C6YGC} & Starting Y grid position \\
\textbf{C1DP} & Precision of the data in bits and data type & \textbf{C4EDC} & Epoch declination [deg] & \textbf{C6YNP} & Number of Y grid points \\
\textbf{C1OBS} & Observer Initials & \textbf{C4EL} & Elevation at C3UT [deg] & \textbf{C7BCV} & Bad channel value \\
\textbf{C1ONA} & Observer name & \textbf{C4EPH} & Epoch of coordinates [years]& \textbf{C7CAL} & Calibration type \\
\textbf{C1PID} & Project ID & \textbf{C4ERA} & Epoch Right Ascension [deg]& \textbf{C7OSN} & Calibration scan/observation number \\
\textbf{C1RCV} & Frontend & \textbf{C4GB} & Galactic Latitude [deg]& \textbf{C7VC} & Velocity correction [km/s]\\
\textbf{C1SNO} & Scan/Observation number & \textbf{C4GL} & Galactic Longitude [deg]& \textbf{C7VR} & Radial velocity of source [km/s]\\
\textbf{C1STC} & Type of observation & \textbf{C4RX} & Reference X position [deg]& \textbf{C7VRD} & Velocity definition code \\
\textbf{C1TEL} & Telescope name & \textbf{C4RY} & Reference Y position [deg]& \textbf{C8AAE} & Aperture efficiency \\
\textbf{C2FL} & EW focus & \textbf{C4SX} & Source X [deg]& \textbf{C8ABE} & Beam efficiency \\
\textbf{C2FR} & Radial focus & \textbf{C4SY} & Source Y [deg]& \textbf{C8EF} & Forward spillover \& scattering efficiency \\
\textbf{C2FV} & NS focus & \textbf{C5AT} & Ambient Temperature [$^\circ$C]& \textbf{C8EL} & Rear spillover \& scattering efficiency \\
\textbf{C2ORI} & Secondary orientation & \textbf{C5DP} & Dew point [$^\circ$C]& \textbf{C8GN} & Antenna gain \\
\textbf{C2XPC} & Az/RA pointing correction [arcsec] & \textbf{C5IR} & Refractive index & \textbf{C11VD} & Phase table names \\
\textbf{C2YPC} & El/Dec pointing correction [arcsec] & \textbf{C5MM} & Atmospheric vapor pressure [mm]& \textbf{C12BW} & Bandwidth \\
\textbf{C3CL} & Length of cycle [sec] & \textbf{C5PRS} & Atmospheric pressure [mm~Hg]& \textbf{C12CF} & Observed frequency [MHz]\\
\textbf{C3DAT} & UT date (YYYY.MMDD format) & \textbf{C5RH} & Relative humidity [\%]& \textbf{C12CT} & Calibration temperature [K]\\
\textbf{C3LST} & LST at start [hours] & \textbf{C6DX} & Delta X offset [arcsec]& \textbf{C12FR} & Frequency resolution [MHz]\\
\textbf{C3NRC} & Number of rx/backend channels & \textbf{C6DY} & Delta Y offset [arcsec]& \textbf{C12RF} & Rest frequency [MHz]\\
\textbf{C3NSV} & Number of switching/phase table variables & \textbf{C6FC} & Reference frame coord code & \textbf{C12RST} & Reference system temperature [K]\\
\textbf{C3PPC} & Number of phases per cycle & \textbf{C6MSA} & Scanning angle [deg]& \textbf{C12RT} & Receiver temperature [K]\\
\textbf{C3SRT} & Integration time [sec] & \textbf{C6NP} & Number of grid points & \textbf{C12SST} & Source system temperature [K]\\
\textbf{C3UT} & UT of observation [hours] & \textbf{C6XGC} & Starting X grid position & \textbf{C12WO} & Water opacity \\
\textbf{C4AZ} & Azimuth at C3UT [deg]  & \textbf{C6XNP} & Number of X grid points &   &   \\
\hline
\end{tabular}
\end{table*}

At the JCMT these GSDD names (known locally as the ``NRAO'' names)
were written to disk files but were mapped to local equivalents in
the acquisition computers. For example \texttt{C12RF}, the rest
frequency, mapped to \texttt{FE\_NUREST} in the acquisition shared
memory system and
was equivalent to the \texttt{RESTFREQ} FITS keyword. A full list of
the equivalences for JCMT can be found elsewhere \citep{SUN229,mtdn85}. As
commissioning took place, new
instrumentation arrived and new facilities were added, the JCMT data
model diverged with many new items being added without consultation
with NRAO. These items are listed in Tables \ref{tab:appa1},
\ref{tab:appa2} and \ref{tab:appa3}.  Class 55 (Inclinometry) is not
included here as the inclinometry data were not archived and therefore
data files describing these observations are extremely rare.

\begin{table*}[!ht]
\caption{JCMT-specific keywords from class 1 to class 4. This includes the three items that were not allocated a class.}
\label{tab:appa1}
\begin{center}
\begin{tabular}{|lp{2.0in}|lp{2.0in}|}
\hline
\textbf{CELL\_V2Y} & Position angle of cell y axis (CCW) & \textbf{C3DASSHFTFRAC} & DAS calibration source for backend calibration (POWER or DATA)\\
\textbf{UAZ} & User az correction & \textbf{C3FLY} & Data taken on the fly or in discrete mode?\\
\textbf{UEL} & User el correction & \textbf{C3FOCUS} & Focus observation?\\
\textbf{C1BTYP} & Type of backend & \textbf{C3INTT} & Scan integration time\\
\textbf{C1FTYP} & Type of frontend & \textbf{C3LSPC} & Number of channels per backend section\\
\textbf{C1HGT} & Height of telescope above sea level & \textbf{C3MAP} & Map observation?\\
\textbf{C1IFS} & Name of the IF device & \textbf{C3MXP} & Maximum number of map points done in a phase\\
\textbf{C1LAT} & Geodetic latitude of telescope (North +ve) & \textbf{C3NCH} & Number of backend output channels\\
\textbf{C1LONG} & Geographical longitude of telescope (West +ve) & \textbf{C3NCI} & Maximum number of cycles in the scan\\
\textbf{C1ONA1} & Name of the support scientist & \textbf{C3NCP} & Total number of xy positions observed during a cycle\\
\textbf{C1ONA2} & Name of the telescope operator & \textbf{C3NCYCLE} & Number of cycles done in the scan\\
\textbf{C1SNA1} & Source name part 1 & \textbf{C3NFOC} & Number of frontend output channels\\
\textbf{C1SNA2} & Source name part 2 or altern. name & \textbf{C3NIS} & Number of scans\\
\textbf{C2PC1} & Angle by which lower axis is north of ideal & \textbf{C3NLOOPS} & Number of scans per observation commanded at observation start\\
\textbf{C2PC2} & Angle by which lower axis is east of ideal & \textbf{C3NMAP} & Number of map points\\
\textbf{C2PC3} & Angle by which upper axis is not perpendicular to lower & \textbf{C3NOIFPBES} & Number of IF inputs to each section (2 for correlator, 1 for AOS)\\
\textbf{C2PC4} & Angle by which beam is not perpendicular to upper axis & \textbf{C3NO\_SCAN\_VARS1} & Number of scan table 1 variables\\
\textbf{C3BEFENULO} & Copy of frontend LO frequency per backend section & \textbf{C3NO\_SCAN\_VARS2} & Number of scan table 2 variables\\
\textbf{C3BEFESB} & Copy of frontend sideband sign per backend section & \textbf{C3NPP} & Number of dimension in the map table\\
\textbf{C3BEINCON} & IF output channels connected to BE input channels & \textbf{C3NRS} & Number of backend sections\\
\textbf{C3BESCONN} & BE input channels connected to this section & \textbf{C3NSAMPLE} & Number of scans done\\
\textbf{C3BESSPEC} & Subsystem nr to which each backend section belongs. & \textbf{C3OVERLAP} & Subband overlap\\
\textbf{C3BETOTIF} & Total IF per backend section & \textbf{C3UT1C} & UT1-UTC correction interpolated from time service telex (in days)\\
\textbf{C3CAL} & Calibration observation? & \textbf{C4AMPL\_EW} & Secondary mirror chopping amplitude parallel to lower axis\\
\textbf{C3CEN} & Centre moves between scans? & \textbf{C4AMPL\_NS} & Secondary mirror chopping amplitude parallel to upper axis\\
\textbf{C3CONFIGNR} & Backend configuration & \textbf{C4AXY} & Angle between cell y axis and x-axis (CCW)\\
\textbf{C3DASCALSRC} & DAS calibration source for backend calibration (POWER or DATA) & \textbf{C4AZERR} & DAZ:Net Az offset at start (inc.tracker ball setting and user correction)\\
\textbf{C3DASOUTPUT} & Description of output in DAS DATA (SPECTRUM, T\_REC, T\_SYS, etc.) & \textbf{C4CECO} & Centre coords. AZ=1; EQ=3; RD=4; RB=6; RJ=7; GA=8\\
\hline
\end{tabular}
\end{center}
\end{table*}
\begin{table*}[!ht]
\caption{JCMT-specific keywords from class 4 to class 7.}
\label{tab:appa2}
\begin{center}
\begin{tabular}{|lp{2.0in}|lp{2.0in}|}
\hline
\textbf{C4DECDATE} & Declination of date & \textbf{C4THROW} & Secondary mirror chop throw\\
\textbf{C4DEL} & Telescope upper axis correction for secondary mirror XYZ & \textbf{C4X} & Secondary mirror absolute X position at observation start\\
\textbf{C4DO1} & Cell x dimension; descriptive origin item 1 & \textbf{C4Y} & Secondary mirror absolute Y position at observation start\\
\textbf{C4DO2} & Cell y dimension; descriptive origin item 2 & \textbf{C4Z} & Secondary mirror absolute Z position at observation start\\
\textbf{C4DO3} & Angle by which the cell x axis is oriented with respect to local vertical & \textbf{C5IR1} & Refraction constant A\\
\textbf{C4EDEC} & Declination of source for EPOCH & \textbf{C5IR2} & Refraction constant B\\
\textbf{C4EDEC2000} & Declination J2000 & \textbf{C5IR3} & Refraction constant C\\
\textbf{C4ELERR} & DEL:Net El offset at start (inc.tracker ball setting and user correction) & \textbf{C6CYCLREV} & Cycle reversal flag\\
\textbf{C4EPT} & Type of epoch, JULIAN, BESSELIAN or APPARENT & \textbf{C6MODE} & Observation mode\\
\textbf{C4EW\_ENCODER} & Secondary mirror ew encoder value & \textbf{C6REV} & Map rows scanned in alternate directions?\\
\textbf{C4EW\_SCALE} & Secondary mirror ew chop scale & \textbf{C6SD} & Map rows are in X (horizontal) or Y(vertical) direction\\
\textbf{C4FRQ} & Secondary mirror chopping period & \textbf{C6ST} & Type of observation\\
\textbf{C4FUN} & Secondary mirror chopping waveform & \textbf{C6XPOS} & In first row x increases (TRUE) or decreases (FALSE)\\
\textbf{C4LSC} & Char. code for local x-y coord.system & \textbf{C6YPOS} & In first row y increases (TRUE) or decreases (FALSE)\\
\textbf{C4MCF} & Centre moving flag (solar system object) & \textbf{C7AP} & Aperture\\
\textbf{C4MOCO} & Mounting of telescope; defined as LOWER/UPPER axes, e.g; AZ/ALT & \textbf{C7FIL} & Filter\\
\textbf{C4NS\_ENCODER} & Secondary mirror ns encoder value & \textbf{C7HP} & FWHM of the beam profile (mean)\\
\textbf{C4NS\_SCALE} & Secondary mirror ns chop scale & \textbf{C7NIF} & Number of IF channels\\
\textbf{C4ODCO} & Units of cell and mapping coordinates;offset definition code & \textbf{C7PHASE} & Lockin phase\\
\textbf{C4OFFS\_EW} & Secondary mirror offset parallel to lower axis (East-West Tilt) & \textbf{C7SEEING} & Seeing at JCMT\\
\textbf{C4OFFS\_NS} & Secondary mirror offset parallel to upper axis (North-South Tilt) & \textbf{C7SEETIME} & SAO seeing time (YYMMDDHHMM)\\
\textbf{C4PER} & Secondary mirror chopping period & \textbf{C7SNSTVTY} & Lockin sensitivity in scale range units\\
\textbf{C4POSANG} & Secondary mirror chop position angle & \textbf{C7SNTVTYRG} & Sensitivity range of lockin\\
\textbf{C4RA2000} & Right ascension J2000 & \textbf{C7SZVRAD} & Number of elements of vradial array\\
\textbf{C4RADATE} & Right Ascension of date & \textbf{C7TAU225} & CSO tau at 225GHz\\
\textbf{C4SM} & Secondary mirror is chopping & \textbf{C7TAURMS} & CSO tau rms\\
\textbf{C4SMCO} & Secondary mirror chopping coordinate system & \textbf{C7TAUTIME} & CSO tau time (YYMMDDHHMM)\\
\hline
\end{tabular}
\end{center}
\end{table*}
\begin{table*}[!ht]
\caption{JCMT-specific keywords from class 7 to class 14. Class 55 is not included here as that data was not archived.}
\label{tab:appa3}
\begin{center}
\begin{tabular}{|lp{2.0in}|lp{2.0in}|}
\hline
\textbf{C7VREF} & Velocity reference code; reference point for telescope \& source velocity & \textbf{C12SCAN\_VARS2} & Names of the cols. of scan table2\\
\textbf{C11PHA} & Phase table: switching scheme dependent & \textbf{C12TAMB} & Ambient load temperature\\
\textbf{C12ALPHA} & Ratio of signal sideband to image sideband sky transmission & \textbf{C12TASKY} & Ratio of signal sideband to image sideband sky transmission\\
\textbf{C12BM} & Correlation bit mode & \textbf{C12TCOLD} & Cold load temperature\\
\textbf{C12CAL} & Units of spectrum data & \textbf{C12TSKY} & Sky temperature at last calibration\\
\textbf{C12CALTASK} & Calibration instrument used (FE, BE, or USER) & \textbf{C12TSKYIM} & Frontend-derived Tsky, image sideband\\
\textbf{C12CALTYPE} & Type of calibration (THREELOADS or TWOLOADS) & \textbf{C12TSYSIM} & Frontend-derived Tsys, image sideband\\
\textbf{C12CM} & Correlation function mode & \textbf{C12TTEL} & Telescope temp. from last skydip\\
\textbf{C12ETASKY} & Sky transmission from last calibration & \textbf{C12VCOLD} & IF V\_COLD\\
\textbf{C12ETASKYIM} & Frontend-derived sky transmission & \textbf{C12VDEF} & Velocity definition code - radio, optical, or relativistic\\
\textbf{C12ETATEL} & Telescope transmission & \textbf{C12VHOT} & IF V\_HOT\\
\textbf{C12GAINS} & Gain value (kelvins per volt or equivalent) & \textbf{C12VREF} & Velocity frame of reference - LSR, Bary-, Helio-, or Geo- centric\\
\textbf{C12GNORM} & Data normalisation factor & \textbf{C12VSKY} & IF V\_SKY\\
\textbf{C12GREC} & Raw data units per Kelvin & \textbf{C13DAT} & Reduced photometric value \emph{or} Spectrum data \emph{or} Reduced data\\
\textbf{C12GS} & Normalizes signal sideband gain & \textbf{C13ERR} & Standard error\\
\textbf{C12INFREQ} & BE input frequencies [GHz] & \textbf{C13RAW\_ERROR} & Raw error is accumulated over the scan, so store at end scan\\
\textbf{C12NOI} & Noise value & \textbf{C13RAW\_ERROR\_OP} & Raw (out of phase) error also to be stored at end scan\\
\textbf{C12REDMODE} & Way of calibrating the data (RATIO or DIFFERENCE) & \textbf{C13RESP} & array of responsivities\\
\textbf{C12SBRAT} & Sideband ratio & \textbf{C13SPV} & Individual beam integrations \emph{or} Raw data\\
\textbf{C12SCAN\_TABLE\_1} & Begin scan table & \textbf{C13SPV\_OP} & Raw out of phase data samples in each phase\\
\textbf{C12SCAN\_TABLE\_2} & End scan table & \textbf{C13STD} & Phase data standard deviation\\
\textbf{C12SCAN\_VARS1} & Names of the cols. of scan table1 & \textbf{C14PHIST} & List of xy offsets for each scan\\
\hline
\end{tabular}
\end{center}
\end{table*}

One feature of the GSDD design was that some classes were
explicitly reserved for local use. Class 9 was used for telescope
dependent parameters and the defined set differed between Green Bank
and the 12m with JCMT adopting a single item, \texttt{C9OT} from the
12m.

The NRAO implementation, not including class 9, includes 26 items not
found in the JCMT version and these are given in
Table~\ref{tab:nraoonly}.
The following list -- which is not intended to be exhaustive -- details
the main discrepancies and major compatibility problems between the NRAO and
GSDD data models. The first part describes in detail the items found
in the NRAO model but not implemented at JCMT:

\begin{table}[t]
\caption{Items of the GSDD model only defined for use at NRAO.}
\label{tab:nraoonly}
\begin{center}
\begin{tabular}{lp{2.5in}}
\hline
\textbf{C1DLN} & Length of Data [bytes]\\
\textbf{C1HLN} & Length of Header [bytes]\\
\textbf{C1SNA} & Source Name\\
\textbf{C2PC} & Pointing Constants(4)\\
\textbf{C2UXP} & User Az/RA Pointing Correction [arcsec]\\
\textbf{C2UYP} & User El/Dec Pointing Correction [arcsec]\\
\textbf{C4DO} & Descriptive Origin(3)\\
\textbf{C4IX} & Indicated X Position [deg]\\
\textbf{C4IY} & Indicated Y Position [deg]\\
\textbf{C6XZ} & X Position at Map Reference Position Zero [deg]\\
\textbf{C6YZ} & Y Position at Map Reference Position Zero [deg]\\
\textbf{C7FW} & Beam Fullwidth at Half Maximum [arcsec]\\
\textbf{C11TP} & Phase Table\\
\textbf{C11VV} & Variable Value\\
\textbf{C12DX} & Delta X\\
\textbf{C12IT} & Total Integration Time [sec]\\
\textbf{C12NI} & Number of Integrations\\
\textbf{C12OO} & O2 Opacity\\
\textbf{C12OT} & O2 Temperature [K]\\
\textbf{C12RMS} & RMS of Mean\\
\textbf{C12RP} & Reference Point Number\\
\textbf{C12SP} & Polarization\\
\textbf{C12SPN} & Starting Point Number\\
\textbf{C12ST} & Source Temperature\\
\textbf{C12WT} & H2O Temperature [K]\\
\textbf{C12X0} & X Value at the Reference Point\\
\hline
\end{tabular}
\end{center}
\end{table}

\begin{description}

\item[\texttt{C1DLN} \texttt{C1HLN}] are not needed at JCMT
  because the length of the header region and the length of the data
  region are encoded in the file format design.

\item[\texttt{C1SNA}] is the source object name and exists as two separate
  items at JCMT, \texttt{C1SNA1} and \texttt{C1SNA2}, to allow the
  object name to be specified in two parts or with an alternative name
  given. \texttt{C1SNA1} is the primary source name and is equivalent
  to the \texttt{OBJECT} FITS keyword. Historically the alternate or
  secondary part of the name was rarely used at JCMT so the name
  change, in hindsight, turned out to be unnecessary.

\item[\texttt{C2PC}] was used at NRAO to specify a four-element
  secondary pointing correction. The JCMT version specifies this as
  four discrete scalar items, \texttt{C2PC1} to \texttt{C2PC4}, rather
  than using an array.

\item[\texttt{C2UXP} \texttt{C2UYP}] are the user Az/RA and El/Dec
  pointing corrections in arcsec but at JCMT these were simply called
  \texttt{UAZ} and \texttt{UEL} with no class prefix and no RA/Dec
  equivalent.

\item[\texttt{C4DO}] is a three-element array labeled ``Descriptive
  Origin'' describing the position and angle of the coordinate system
  defined by the observer. At JCMT this was implemented as three
  distinct items \texttt{C4DO1} through \texttt{C4DO3} and specified
  the observing cell size and position angle with respect to local
  vertical. There was disagreement between NRAO and JCMT on the
  definition here as the three elements at NRAO referred to the
  horizontal and vertical position and the position angle with respect
  to the horizontal axis. Documents and source code from JCMT indicate
  these items were not used and are duplicates of items \texttt{C6DX},
  \texttt{C6DY} and \texttt{C6MSA}.

\item[\texttt{C4IX} \texttt{C4IY}] are the coordinates of the
  telescope as measured by the encoders. This information was not
  recorded by JCMT.

\item[\texttt{C6XZ} \texttt{C6YZ}] specify the position of the map
  origin. These coordinates are not stored at JCMT as the map is
  defined in terms of offsets from the specified tracking centre.

\item[\texttt{C7FW}] is the beam full width at half maximum in arcsec
  at NRAO but at JCMT the item used is \texttt{C7HP} and most JCMT data
  files do not seem to set it.

\item[\texttt{C11*}]
  Most sub-mm telescope divide an ``observation cycle'' into a series
  of ``phases'', where the separate phases represent different states
  (for example, on-source, off-source, cal-diode on, cal-diode
  off). The relevant information is stoed in class 11, the ``phase
  block''.  At JCMT \texttt{C11VD} specifies the names of the columns
  of the phase table information stored in \texttt{C11PHA} where the
  dimensionality is specified by \texttt{C3NSV} (number of phase table
  variables) and \texttt{C3PPC} (number of phases per cycle). NRAO use
  \texttt{C11VV} to store the values of a single switch state and the
  phase table is \texttt{C11PHT}.

\item[\texttt{C12IT}] is the total time spent collecting data,
  including any blanking time.  This item was not used at JCMT.

\item[\texttt{C12NI} \texttt{C12SPN}] indicate the number of integrations (or
  channels for spectral line data) and the starting point (channel) in the data vector.
  UniPOPS used this information to limit display and processing to a sub-set of the
  data array and to associate those limits with the data on disk.
  JCMT data did not need these quantities for any
  similar purpose.

\item[\texttt{C12ST} \texttt{C12RMS}] are the computed source
  temperature and the RMS value. The JCMT online observing system did
  not calculate these.

\item[\texttt{C12SP}] is a description of the polarization type and
  angle encoded in an eight character field. This item was not used at
  JCMT.

\item[\texttt{C12RP} \texttt{C12X0} \texttt{C12DX}] These give the
  reference channel, X value at the reference channel, and spacing along
  the X axis.  For spectral line data, the X axis is velocity at
  each channel and for continuum data this is position along the
  direction of telescope motion for each continuum integration in
  that scan.  These items were not used at JCMT.

\item[\texttt{C12WT}] is the water temperature. Not measured directly
  at JCMT during this period.

\item[\texttt{C12OO} \texttt{C12OT}] is the oxygen opacity and
  temperature. Not measured at JCMT.

\end{description}

The following items are found in both implementations.
Some discrepancies are noted as follows.

\begin{description}

\item[\texttt{C1DP}] This is used to specify the precision and data
  type used to store the instrument data. This was used in early
  variants of the JCMT system but was later dropped due to the data
  format being able to report the data type associated with each item.

\item[\texttt{C1ONA}] JCMT used this as a synonym for \texttt{C1OBS}
  and instead added \texttt{C1ONA1} to indicate the name of the
  support scientist for the observing run and \texttt{C1ONA2} to
  indicate the name of the telescope operator.

\item[\texttt{C1STC}] specifies the type of observation. At NRAO this
  was defined as two 4 character strings defining the type of data and
  the observing mode. For example \texttt{LINEPSSW} for a
  position-switched spectral line observation. JCMT used this item
  solely to define the switching mode (position-switched, beam switch,
  frequency switch and no switch), preferring instead to use
  \texttt{C1FTYP} to specify the frontend type (heterodyne versus
  bolometer) and \texttt{C1BTYP} to indicate the backend type (line
  versus continuum). In
  later versions JCMT dropped \texttt{C1STC} completely, preferring to
  specify the switching mode explicitly in \texttt{C6MODE}.

\item[\texttt{C3UT}] is the Universal Time in decimal hours, yet at
  JCMT it was decided that this should refer to UT1.

\item[\texttt{C4CSC}] The JCMT coordinate system codes \citep{mtdn12}
  were a two-character code such as RB to indicate B1950 RA/Dec.
  NRAO used a completely distinct set of codes using eight
  characters; RB being equivalent to 1950RADC. The full list is shown
  in Table~\ref{tab:coordcodes}.

\begin{table}
\caption{Coordinate codes used at NRAO and JCMT for item
  \texttt{C4CSC}.}
\label{tab:coordcodes}
\begin{center}
\begin{tabular}{lll}
\hline
 & NRAO & JCMT \\ \hline
Galactic & GALACTIC & GA \\
B1950 RA/Dec & 1950RADC & RB \\
Epoch RA/Dec & EPOCRADC & RD \\
Mean RA/Dec & MEANRADC & -- \\
Apparent RA/Dec & APPRADC & -- \\
Apparent HA/Dec & APPHADC & EQ \\
1950 Ecliptic & 1950ECL & EC \\
Epoch ecliptic & EPOCECL & -- \\
Apparent ecliptic & APPECL & -- \\
Azimuth/Elevation & AZEL & AZ \\
User defined & USERDEF & UD \\
J2000 RA/Dec & 2000RADC & RJ \\
Indicated Ra/Dec & INDRADC & -- \\
\hline
\end{tabular}
\end{center}
\end{table}

\item[\texttt{C5IR}] Whilst JCMT did use \texttt{C5IR} to report the
  mean refractive index, the JCMT implementation also stored the three
  refraction constants defined in the JCMT refraction model
  \citep{mtin26} as \texttt{C5IR1}, \texttt{C5IR2} and \texttt{C5IR3}.

\item[\texttt{C6FC}] is the coordinate frame to use when offsetting,
  which allows the offset system to be distinct from the telescope
  tracking centre. At NRAO this was an eight character string made up
  of two four character components (polar versus cartesian and step
  versus scanning). At JCMT this item was an integer indicating which
  coordinate frame should be used with options of AZ=1, EQ=3, RD=4,
  RB=6, RJ=7 and GA=8 (using the same definition explained in item
  \texttt{C4CSC}).

\item[\texttt{C7VRD}] is defined as the velocity definition and
  reference at NRAO, by combining two four character strings into a
  single item. It describes how the source radial velocity,
  \texttt{C7VRD}, should be intrepreted. The allowed velocity
  definitions were RADI (radio), OPTL (optical) and RELV
  (relativistic). The velocity reference was allowed to be LSR (Local
  Standard of Rest), HELO (Heliocentric), EART (earth), BARI
  (barycentre) and OBS (observer). At JCMT this item was reserved
  entirely for the velocity definition but deprecated in later
  versions. The velocity definition was later defined in
  \texttt{C12VDEF} (allowed values being RADIO, OPTICAL and
  RELATIVISTIC) and the standard of rest indicated in \texttt{C12VREF}
  (allowed values being TOPO(centric), LSR, HELI(ocentric),
  GEO(centric), BARY(centric) and TELL(uric)).

\end{description}

\section{National Radio Astronomy Observatory}

The 12-m Telescope was upgraded to write GSDD format data in the
summer of 1986 \citep{tcus23,1987NRAO30}; requiring that the data
analysis system was also updated to understand it.

In 1988 the NRAO decided for a number of reasons to unify the data
reduction systems for its single-dish telescopes: the Tucson 12-m, and
the Green Bank 300\,ft and 140\,ft telescopes.  At the time all three
telescopes used what looked like a very similar data reduction system,
the People Oriented Parsing Service \citep[POPS;][]{1982POPS}
But, at the code level the applications in Green Bank
and Tucson had been diverging rapidly since the early 1980's,
essentially due to the different computer architectures at the two
sites (early 1970's Modcomps in Green Bank and mid-1980 DEC VAX's in
Tucson).  The NRAO wanted to reduce maintenance costs as different
staff were needed to maintain and develop each version.  The NRAO was
also migrating to Unix-based (primarily Sun) computers, a change that
would require major modifications to POPS. The unified analysis system, UniPOPS
\citep[][\ascl{1503.007}]{UNIPOPS}, was started in early 1989 and first released to users in early
1991 \citep{1991BAAS...23..535V}.  Although the 300\,ft collapsed in
1988 \citep{1990BAAS...22..487V}, and the 140\,ft was
decommissioned for routine general-user astronomy in 1999, UniPOPS
is still in use today at
some level by the University of Arizona who took over the running of
the 12-m telescope in 2004.

Since the majority of the FORTRAN code that was modified to create
UniPOPS came from the 12-m version of POPS, the UniPOPS developers
decided that UniPOPS would also inherit with little modification the
underlying data structure and export formats of the 12-m version of
POPS.  Internally, the UniPOPS data structure used to hold the data
is the same as the 12-m version of the GSDD data model with additional
items added as described in section~\ref{sec:datamodel} and a new Class 9 to hold
the values that were unique to to the Green Bank telescopes. The UniPOPS
file format is nearly identical to the POPS Data File (PDFL) format
in use at the 12-m prior to UniPOPS. This
data structure and file format were used by UniPOPS to hold data at
all stages of processing (raw, calibated, averaged, smoothed, etc.).
Adapting 140\,ft and 300\,ft data to use the GSDD data model was
relatively easy, good evidence that GSDD was indeed a rather versatile
and useful standard.  Additional details on the export file format
used by UniPOPS and the format of the raw data written at each
NRAO telescope are provided in the next section.

Two modifications were made to the PDFL files when they were
incorporated into UniPOPS, solely to boost the performance of the
system.  The binary representation was changed from that of the DEC
architecture to that of Sun workstations.  And, the index that was at
the start of a PDFL file was extended to include such items as the sky
location and observing frequency to expand the items that could be
efficiently searched in UniPOPS.  To distinguish the UniPOPS
Sun-specific exported files from VAX PDFL files, the NRAO developers
changed the name of the export format to Single Dish Data (SDD) format.
Other than a modification that expanded the capabilities of the index
section of the NRAO SDD files, the SDD format adopted for UniPOPS
(Fig.~\ref{fig:nraosdd}) remained unchanged until UniPOPS was retired
at the NRAO in the mid-2000's.

By the late 1980's, users of the NRAO telescopes were very interested
in seeing a FITS format implemented for the NRAO's single-dish
telescopes (see \S\,\ref{sec:sdfits}). By the mid 1990's,
UniPOPS could export and import data in Single Dish FITS (SDFITS) and
SDD formats, as well as many
of the historical NRAO formats.  The NRAO found that very few users
went away with SDFITs format; most took home SDD files.  Since
users were installing UniPOPs on their home computers, they probably
found transporting SDD files more convenient than using
SDFITS files.  It was probably very rare that a UniPOPS SDD file was
imported into another analysis system.  For example, a separate
utility was developed that would prepare data files that could be
imported into the \textsc{class} package.
Furthermore, when SDFITs was released, few FITS readers at the
time could actually usefully import binary tables.  Thus, we suspect
that frequent observers grew into the habit of avoiding SDFITS files.

\subsection{SDD File Format}

\begin{figure}[t]
\begin{center}
\includegraphics[width=0.5\columnwidth]{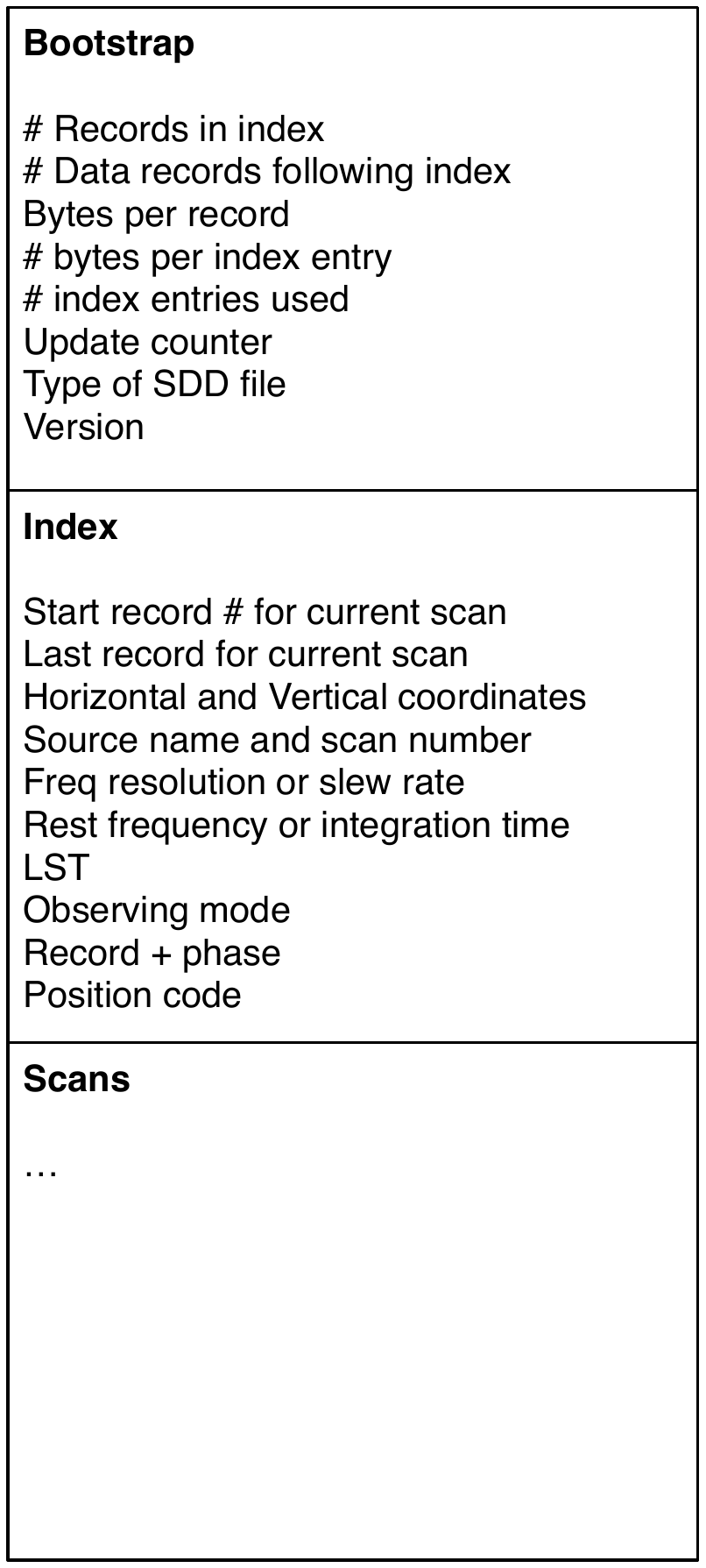}
\end{center}
\caption{Layout of a UniPOPS data file. The concepts are similar to
  those used in the GSD format (Fig.~\ref{fig:jcmtgsd}). The bootstrap
  field describes the basic layout of the file and the index indicates
  where each of the scans are located in the file. A key difference
  between GSD and SDD is that GSD contains a single observation
  whereas an SDD file contains many observations for a single science
  program.}
\label{fig:nraosdd}
\end{figure}

The layout of an SDD file is shown in Fig.~\ref{fig:nraosdd}
\citep[see also][]{UNIPOPS}.  The file consists of 3 parts: a
bootstrap record, the index, and the data.  An SDD file has an integer
number of records where the size of a record is given in the
bootstrap.  Each section (index and data) is also an integer number of
records.  Within the data section, each individual ``scan'' is one
instance of the data structure that evolved from the original GSDD
data model.  Each scan occupies an integer number of records within
the file (any extra space is padded with zeros).  An SDD file can hold
both spectral line and continuum data.  The type of data is indicated
by the \texttt{C1STC} value found in the data for each scan. The order
and type of the values in the bootstrap record and within each index
entry are set by the version number recorded in the bootstrap.
The \texttt{update counter} item in the bootstrap record was an integer that
was incremented each time the file was modified.  This was used at the
12-m where the telescope control system was writing to multiple SDD files
while one or more running UniPOPS sessions were reading from the same
set of SDD files.  This is not a self-describing format like FITS.

The bootstrap record contains information used to read the index
records.  The size of the index entry and record were chosen so that
there are an integer number of index entries in each record without
any extra space.  The index record is read into memory in UniPOPS when
an SDD file is opened and the copy in memory is kept in sync with the
contents of the file as changes are made.  Data selection in UniPOPS
only uses the fields in the index.  Every scan in the file must have
an entry in the index section.  Empty index entries have zeros for the
start and last record numbers.  The current largest index number in
use is indicated in the bootstrap.  If the data query involved the
index associated with one of the SDD files being written by the
12-m during observing, then the \texttt{update counter} value in
the bootstrap record on disk was checked.  When a change was seen in that
value then the copy of the index in memory was regenerated from the
data file before the data query was done.

Each index entry indicates where the associated scan data starts and
ends.  The scan data consists of a preamble, which is 16 short
integers giving the number of classes and the starting 8-byte location
in the header where each class of header words started.  Within each
class, the type and order of each value is fixed.  The \texttt{C1HLN}
value gives the length of all of the header values.  Over time, new
items were added to the ends of some classes so that UniPOPS retained
the ability to read previous versions of the SDD format by not
attempting to read header words past the end of a class as indicated
by the values in the preamble and \texttt{C1HLN}.  These new items are the
differences between the NRAO and JCMT versions of the GSDD data model
mentioned in section~\ref{sec:datamodel}.  Those differences grew over time
as needed by NRAO to accomodate new instruments, observing techniques and
reduction methods. All numerical
values in the header are stored as 8-byte floats.  All string values
are stored as multiples of 8 characters, depending on the specific
value.  The data vector immediately follows the header.  The data are
always 4-byte floats.

In order to accomodate spectral line and continuum data within the
same structure and minimize the amount of space needed to store the
associated header, some header values have 2 meanings depending on the
type of data.  This is most obvious in class 12, where the associated
X axis is described.  For spectral line data the X axis is the
frequency or velocity at each channel.  For
continuum data, the data vector is a series of regularly sampled
data (each sample is one integration) so the X axis is related to
the position on the sky as the telescope is slewed.  UniPOPS was
started either in spectral line mode or continuum mode and would need
to be restarted to switch modes.  It was never possible to work on
both continuum data and spectral line data within the same session of
UniPOPS so typically a single SDD file only contained one type of
data although that was not required by the file format.

\subsection{SDD Usage}

UniPOPS dealt directly only with SDD format files.  An SDD file could
contain raw data, individual integrations and calibrated data. When
writing to an SDD file, UniPOPS could extend that file by appending to
the end or it could overwrite existing data in the file provided that
the size of the scan being overwritten was as least as large as the
scan being written.  In either case, the appropriate index entried was
updated.  In the case of appending to the file, the next index
location after the current end as indicated in the bootstrap record
was used.  The index entries do not need to reflect the order that the
data appear in the file, although that typically is the case.  If a
user tried to overwrite a scan with a scan with more channels UniPOPS
would append the new scan to the file, replace the index entry for the
original scan with an appropriate index entry for the new scan, and
replace the original scan records in the SDD file with zeros.

UniPOPS provided observers with access to their raw data in near
real-time.  For the 12-m this was direct access to the current set of
SDD files being written by the telescope control software.  Multiple
files could be written at the same time, depending on the backend, and
UniPOPS could access the desired data from any of those files
while the data was being taken.  For long observing sessions multiple
versions of each backend-specific file were written. The 12-m also provided
SDD files containing system temperature across the bandpass for each
scan.  UniPOPS provided separate methods for accessing that calibration
data but the file format was identical to all other SDD files.  For
the 140\, ft, the raw data was written in the original telescope
format produced by the Modcomps.  This raw 140\, ft telescope format
predates the GSDD data model.  A conversion step to the NRAO version of
the GSDD data model was necessary for UniPOPS to use that data.
While observing, that conversion step happened on demand within UniPOPS.
Access to the raw 140\, ft data within UniPOPS could be done remotely
by an observer running UniPOPS at their home institution.  The
UniPOPS user could then choose to save that raw 140\, ft data directly
to disk in an SDD format file or they could process the data and only
save those scans to disk.  A separate data conversion tool was also
provided to convert an entire observing session at the 140\, ft from
raw telescope format data to an SDD format file which could be
read directly by UniPOPS without any network connection to the
raw data.

SDD format files could be used interchangeably for input and output by
UniPOPS.  Typically a raw, uncalibrated data set was used as input and
the user would save processed spectra to a separate SDD file.  UniPOPS
users could choose to save their data to disk at any stage of processing.
Any single SDD file could contain raw, calibrated, or reduced data in
any combination.  Typically most users kept the raw data separate from
the processed data as that made it simpler to keep track of what
had been done. A single output SDD file often contained the same data
at different processing steps.  The UniPOPS user needed to keep track
of what had been done to the data as no processing history information
was associated with data either internally or in the SDD file.

With the interactive UniPOPS environment, users had the ability to modify
any of the GSDD data model items (header values) for any scan.  These
header values were referenced by the UniPOPS interpreter using slightly
more readable names (e.g. \texttt{C1SNA} is \texttt{OBJECT}
in UniPOPS). The UniPOPS Cookbook \citep[]{UNIPOPS} uses those more
readable names to reference the GSDD data model items.  Internally,
the compiled code that comprises UniPOPS (mostly fortran) uses the
original GSDD data model names (known at the JCMT as the ``NRAO'' names).

The number of scans that an SDD file can contain is set
by the size of the index section.  Scripts were provided with UniPOPS
to expand an existing SDD file if more index space was necessary.
UniPOPS could not read or write SDFITS directly. Separate conversion tools
were necessary to produce and consume SDFITS.  Conversion tools were also
provided for historical NRAO formats including the PDFL format
used at the 12-m prior to UniPOPS.

Archives from both the Green Bank 140\, ft and Tucson 12-m telescopes
exist.  For the 12-m, there are about 200\, GB of archived SDD format
files.  The archive from the 140\, ft consists entirely of telescope
format files.  The current Green Bank single dish analysis package,
GBTIDL \citep[][\ascl{1303.019}]{2006ASPC..351..512M}), can read
archived SDD files.  GBTIDL uses SDFITS as it's primary data format.

\section{James Clerk Maxwell Telescope}

\subsection{Requirements}

During the development of the JCMT software libraries at the Mullard
Radio Astronomy Observatory, a number of options were considered for
the raw data file format.  Two obvious options were available in the
astronomical community in the form of the Flexible Image Transport System
\citep[FITS;][]{1981A&AS...44..363W} and the Starlink Hierarchical
Data System \citep[HDS;][\ascl{1502.009}]{1982QJRAS..23..485D,2015HDS}.

FITS was discounted as the primary data format because of the large
amount of overhead required to format the header information when
writing files and the inability of the format (at that time) to store
more than one data array or table in a file. FITS files at the time
were not capable of storing binary tables and ASCII tables were all
that was possible \citep{1988A&AS...73..365H} and those were not
standardised until 1987. It was also felt that the DEC Backup Utility
was more reliable for transport and archiving than using a specialist
FITS tape format. Whilst the FITS community would eventually support
multiple data arrays \citep{1988A&AS...73..359G} and binary tables
\citep{1995A&AS..113..159C}, it was not possible to wait for that to
happen.

HDS was discarded for I/O efficiency reasons and the inability for the
entire file to be mapped into memory in one operation. Additionally it
was felt that the HDS library API required too many calls to do simple
tasks, and although these calls could be wrapped in higher level
subroutines, the overhead associated with the many lower level calls
would be too high. One further option was to use the NRAO 12-m file format (PDFL)
but that also suffered
(from the JCMT perspective) from serious I/O issues and could not be used
on the acquisition hardware initially targeted for JCMT.

The computer used during testing and commissioning in 1985/86 was a
VAX 11/730 with 4\,MB of RAM and which had severe performance
limitations. This was upgraded to a MicroVAX with 16\,MB of RAM just
before operations started at JCMT in 1987 but performance was the key
design driver: the control system was required to minimize the
overheads in data capture and therefore maximize the observing
time. The VAX Record Management System (RMS) was the basis of
all standard VAX records-based file handling. The performance of
this system was not suitable for real-time operation
as it was not acceptable for the system to pause while
opening or closing or extending a file in the middle of the data
collection. Furthermore, limits on the maximum record length in RMS
meant that additional complexity would be required when writing out
data from long observations. The JCMT disk I/O approach was instead
designed to utilize the VAX System Services library that allowed a
program to map a section of virtual memory
(referred to as a Global Section) and then manage the scalar
and array data in that memory directly in the program. This was very
fast and did not cause the problems encountered with RMS. Performance
benchmarks on a VAX 750 \citep{mtin33} suggested that I/O operations
using RMS were approximately five times slower than using a Global
Section. The use of a Global Section also allowed other applications
read access to the contents of the file whilst it was being
written and also meant that the data already acquired would be usable
even if the acqusition software crashed mid-observation.

These requirements led to a new disk format being devised and an associated
I/O library written which used the GSDD data model, but used Global Sections
for writing to disk. This led to the JCMT implementation of the library being
known as the Global Section Datafile System
\citep[GSD;][]{mtin33}\footnote{In retrospect, the similarity of
  acronyms between GSD and GSDD -- two quite separate concepts -- was
  rather unfortunate. The naming of the library as GSD eventually led
  to JCMT users referring to the files as being of ``GSD format'' and
  it being assumed that ``GSDD format'' was an historical artefact.}.
The file format
design was influenced by the NRAO idea of a self-describing GSDD
implementation and also the concept of an ``in memory data base
management system''\footnote{A database system designed to work
  entirely in memory rather than requiring lots of disk I/O. See also
  \url{http://en.wikipedia.org/wiki/In-memory_database}.} from the
\texttt{MON} library being used in the JCMT control
system.\footnote{The \texttt{MON} library was a shared memory system,
  based on Global Sections, in use at the JCMT to allow the individual
  control system tasks to easily share state information. It was the
  precursor to the Noticeboard System \citep[NBS;][]{SUN77}.}  JCMT
adopted the GSDD data model in the hope that downstream the data
reduction systems could be compatible through the shared metadata
conventions.

Unlike the NRAO PDFL/SDD files which grow throughout the night as more
data are taken, a JCMT GSD file was only required to store data from a
single observation. At JCMT an observation was defined as data being
taken in a single switching mode at a single tracking position with a
single instrument frontend/backend combination. A single observation
could include multiple offsets in a grid or on-the-fly map and
includes the full map area, rather than a single row or column. This
approach resulted in more files to track in a night but was felt to
simplify the acqusition software (each observation was completely
independent of what had gone before), and make it easier to distribute
subsets of a night's data amongst different observers (a pre-requisite
for flexible scheduling) and simplify queries for individual
observations from the data archive. Of course, this meant that the
data reduction packages had to do more work to collate related
observations into a coherent data set as they now worked with many
independent files rather than being able to treat a night's observing
as a single coherent entity.

\subsection{File Format Design}

\begin{figure}[t]
\begin{center}
\includegraphics[width=0.5\columnwidth]{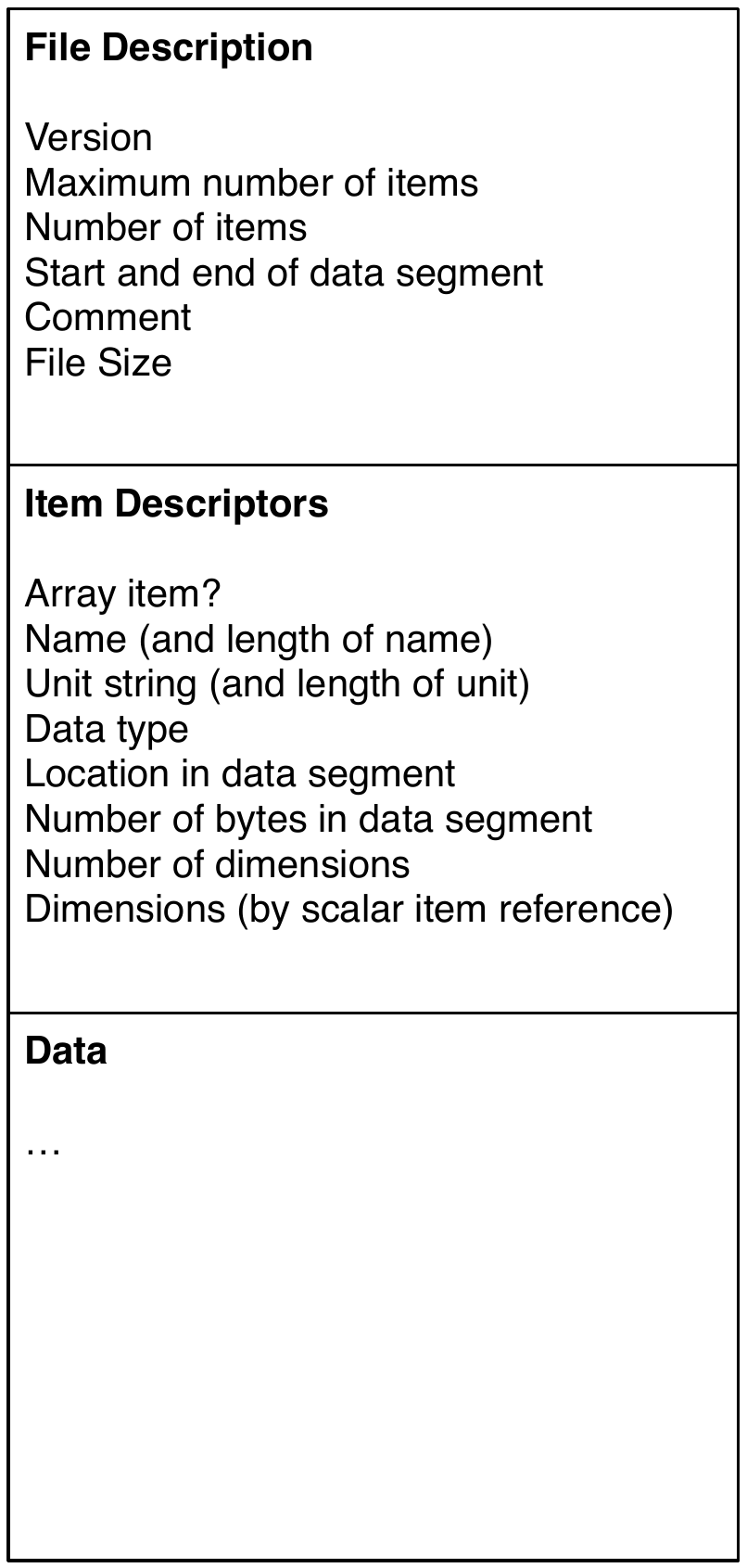}
\end{center}
\caption{Layout of a JCMT GSD data file. The file descriptor indicates
where the data starts and the number of items in the data. The item
descriptors describe each of those items and where they are located in
the data segment. The size of each dimension in array items is defined
in terms of other scalar items. The file was pre-allocated by the
acquistion system at the start of the observation rather than being
continually extended.}
\label{fig:jcmtgsd}
\end{figure}

The layout of a JCMT GSD format file is shown in
Fig.~\ref{fig:jcmtgsd} \citep[see also][]{mtdn84}. The file is split
into three segments: the file descriptor, the item descriptors and the
data itself. The file descriptor contains a general description of the
file indicating its version, the number of items written and the start
position of the data array. The item descriptors define each of the
items in terms of the label and units and the position within the data
array. The data itself is a single block at the end of the file
following the item descriptions; the item descriptions having defined
exactly where in the data array a relevant item is located and how
many bytes in the data array it occupied.

For array items (GSD supported up to 5 dimensions), the
identity of each dimension is specified in terms of the number of a
scalar item. This allows the label and unit to be associated with each
dimension of an array item in addition to the size of the dimension. A
negative number of dimensions indicates that an item is a scalar that
defines an array dimension. For example, the \texttt{C11PHA} array
entry in a JCMT DAS spectrum \citep{1986SPIE..598..134B} is
dimensioned according to the scalar items \texttt{C3NSV}, the number
of phase table variables, and \texttt{C3PPC}, the number of phases per
cycle. The item descriptor for \texttt{C11PHA} would therefore contain
a \texttt{dimensions} array of two elements containing the item
numbers (position in the item descriptor section) for \texttt{C3PPC}
and \texttt{C11PHA}. A library user would then look up those two items
to determine the dimensionality of \texttt{C11PHA}.

This file design resulted in a fully self-describing system
where there was no requirement for items to be grouped by class in the
file and no requirement for the order of items to be pre-determined
(an issue for the NRAO implementation where the order was specified in
a compiled include file requiring that the order of items within a
class be preserved and also that new items could only be added to the
end of a class). A user of the format could either
request an item by number or request an item by name. Storing the
units with the data also allowed for more flexibility in data model
representation at the expense of more logic in the application code
that might have to understand unit conversions. Application software
would use the file version number to decide which variant of a data
model was present in the file. At JCMT this became important as the
system evolved in the first few years.\footnote{Version 4 of the JCMT
  data model was the first stable implementation, released late in
  1988, and the final version was release 5.3. It is version 5.3 that
  is documented here.}

The JCMT format implementing GSDD supports the standard Fortran data
types of byte, word, logical, integer, real, double and character
strings, and uses VAX floating point format \citep[see][for more
information on VAX floating point
format]{Payne:1980:VFP:641845.641849}. To simplify the format,
character strings have a fixed size of 16 characters, item names are
fixed at 15 characters and unit strings are fixed at 10
characters. The format supported the concept of a ``null'' value by
reserving the most negative value of each data type for that purpose
(using a single space as the null character value and false as the
null logical value). Additionally, the JCMT GSD library supported data
type conversion, allowing a user to request a value in a different
type to how it was stored natively in the file. This was an important
aspect of the library interface, simplifying code required by the
reduction software, enabling users of the library to request data in
the form most suitable for them. This feature was influenced by
earlier work on the Starlink Catalog Access and Reporting (SCAR)
relational database management system for astronomical catalog
handling \citep{SUN70}.\footnote{HDS also supported automatic type
  conversion \citep{SSN27}. The authors are not sure when equivalent
  facilities were added to FITS I/O libraries.}

\subsection{Format Usage}

The JCMT took data in the GSD format for all instruments (heterodyne
and continuum) from the telescope commissioning (circa 1986) to the
delivery of SCUBA in 1996 \citep{1999MNRAS.303..659H}. The GSD format
continued to be used for heterodyne instruments until the delivery of
the new ACSIS correlator in 2006 \citep{2009MNRAS.399.1026B}.  SCUBA
and newer instruments wrote data in the Starlink extensible
\emph{N}-dimensional Data Format \citep[NDF;][]{2015NDF}, although
SCUBA's data model was not precisely copied for ACSIS and
SCUBA-2 \citep{2013MNRAS.430.2513H} data. NDF had a key advantage that
it was being used throughout the Starlink Software Collection as the
primary data format \citep{1992ASPC...25..126A}. Writing data using
NDF meant that JCMT data files had immediate access to all the
visualization and analysis applications already available to the
community such as KAPPA \citep[][\ascl{1403.022}]{SUN95}. Many of the
performance worries from the mid-1980s concerning the overhead
associated with the HDS library were no longer relevant in the late 1990s.

The GSD data access library was a VAX-specific library
\citep{1986QJRAS..27..675.,mtdn84} written in Fortran and making
extensive use of VAX system calls. When the last instrument moved off
of the VAX/VMS data acquisitions computers the format could no longer
be used and was retired. There was little motivation to port the data
model to the newer instruments as it was clear by this time that GSDD
had not succeeded and that NDF would be more useful to the JCMT user
community despite the resulting necessity for new ways of describing raw JCMT
data. There was seen to be no advantage to moving the GSDD class and
item names to the newer NDF-based raw data models. Indeed, as
described in sec~\ref{sec:hidden} the standards effort was dead and it
was not obvious to later
users and software developers from where such opaque names had originated.

The GSDD data files are archived at the Canadian Astronomy Data
Centre and approximately 440\,000 GSD format files are in the
archive, totalling approximately 30\,GB. In order to access these data
files on a Unix system a new read-only version of the GSD library was
written in C \citep[][\ascl{1503.009}]{SUN229} and integrated into the standard data
reduction tools SPECX \citep[][\ascl{1310.008}]{SPECX,1990JCMTP...9...25P}, COADD
\citep[][\ascl{1411.020}]{COADD}  and JCMTDR
\citep[][\ascl{1406.019}]{SUN132}.  The GSD format is relatively
simple and the main complication in the new C (and later pure Java)
implementations was the conversion of VAX floating point format to
IEEE format. Furthermore, computers were sufficiently more powerful
by the time the Unix version was written that there was no need to use
memory mapping; the entire contents of a file is read into memory.
GSD was solely used as a data acquisition format at JCMT, with there
being one application on the VAX to enable the editing of contents if
there was a need to fix some metadata. Data reduction applications
never wrote data out in GSD format and the Unix port of the library
did not have the ability to write a GSD file.
A Perl interface to the Unix C GSD library \citep{1999ASPC..172..494J}
was implemented to allow the preview of spectra for remote observers
when doing flexible scheduling \citep{1997ASPC..125..401J}.

The GSD format files are no longer part of the publically available
query system at the CADC. This was driven by funding constraints when
the CADC system was re-engineered to use a common internal data model
\citep{2013ASPC..475..159R} and a requirement that federal interfaces
be compliant with Canadian language regulations. The JCMT Science
Archive \citep[JSA;][]{2015Economou} therefore does not contain GSD
data.  To extend the useful life of the GSD format observations and to
make the observations available to the widest possible community
through the JSA and the Virtual Observatory, there was a project to
convert the GSD heterodyne files archived at CADC to the modern ACSIS
format \citep{OCS_ICD_022} such that they can be processed (baseline
subtracted, co-added, placed into data cubes) using the
standard JCMT data reduction pipelines
\citep{2015ACSISDR,2008ASPC..394..565J}. The SMURF data reduction
application (\ascl{1310.007}\nocite{2013ascl.soft10007J}) contains the
ability to read GSD files and migrate them to the modern format
\citep{SUN259}.  The GSD files from the earlier continuum instruments,
such as UKT14 \citep{1990MNRAS.243..126D}, will remain in the archive
although they will not be visible through the JSA interface.

\section{Retrospective}

GSDD has had a mixed history and in this section we look back on the
good and bad of GSDD.

\subsection{The hidden standard}
\label{sec:hidden}

The key failure of GSDD was that most of the
developers and users of the format did
not realize that it was a standard and therefore there was no impetus
for the respective observatory staff to continue to communicate as
systems evolved. The initial developers of the JCMT system did not
maintain the data acquisition software in Hawaii and, at
NRAO, the lead developer of the 12m GSDD system left NRAO before the
end of the 1980s.  Interviewing staff from NRAO and JCMT following
the respective implementations of GSDD compatible systems, it was very
rare for anyone to remember that there was an intent for a standard to
be in place. As can be seen from the evolution of the JCMT class names
and the divergence of data models, items were added to the respective
data formats without any communication between the nominal GSDD
partners. 12-m development continued with tweaking of the acquisition
and reduction formats independently.  As the GSDD model evolved, the
NRAO implementation resulted in 24 items that are not present in the
JCMT implementation (not including the classes explicitly specified to
be locally defined), and 154 items that are defined by JCMT but not
defined by NRAO.

The goal of unified data reduction software understanding GSDD never
materialized. Indeed, interoperability usually occurred, if at all, by exporting
the files into a completely different format that could be understood
by \textsc{class}.

In conclusion, it is impossible for a standard to survive as a standard
if no-one knows they are using a standard; the effort must be made
to broadcast and properly document the effort within the wider
community as part of the original development.

\subsection{A model must define the values and units}

Whilst the data model provided a reasonable baseline for how to name
items, it broke down almost immediately when it came to storing values
in those items. For example, the coordinate codes, \texttt{C4CSC},
were not standardised, the reference frame coordinate code,
\texttt{C6FC}, had a whole different concept at JCMT and NRAO and,
indeed, the specification of how observing grids were defined at both
observatories differed despite sharing the same underlying item
names. If an attempt had ever been made to transfer data between
observatories special code would have to be written to import the
data, removing most of the gains of a shared model.
The was due to a failure to fully develop the standard prior to
starting its implementation. In some sense, the development/initiative
was not initiated with/subjected to proper project management
procedures as we currently understand them.

\subsection{Embrace Flexibility}

A major advantage of GSDD is that the standard actually allowed sites
to alter the format and data model as they saw fit.
NRAO sites using the NRAO file format had to follow some minor rules in order to
guarantee that any other site's GSDD reader could still manage the
files.  Such rules as: do not touch the pre-defined keywords (which
were to have predefined byte sizes and were always to be in a certain
order at the start of a class), you are free to add new keywords to
any class but only at the end of the pre-defined section of each
class, modify class 9 for your particular telescope, modify class 10
as convenient, and be sure to use the well-defined pre-amble to
designate the byte at which every class begins.  We maintain that GSDD
was actually a very good implementation for its time because these
rules could be easily adhered to while simultaneously giving
sufficient versatility to each telescope. The JCMT GSD file format
encouraged far more flexibility than this since the constraints on
class keyword ordering were removed and software did not need to
compile-in knowledge of where the individual items were meant to be
located in the file. This led to much more explosive and
dynamic modifications to the data model in the early years of
telescope operations.

\subsection{Too much flexibility is not always good}

The alternative view is that allowing a class 9 for particular
telescopes to use as they liked was an impediment to
standardization. In many cases an item being added to class 9 could have been
made generically useful with some discussion or may well have been very
similar to an item already in use by another telescope. The use of the
escape hatch class should have been treated as a last resort after debate
within the community. Only when it was determined that a
particular item was unique for a telescope should class 9 have been used, and
even then a case could be made that it would still be more helpful for
the item to have been placed in the correct class and documented as such, to
help the next telescope that required similar functionality. In some
sense this was the approach used at JCMT (without the communication
effort) which was simply to ignore class 9 completely and add items to
the ``correct'' classes without discussion in the wider
community. As the JCMT model evolved it was soon clear that many of
the items were not relevant to particular observing modes. Rather than
attempting to always write them out regardless, it was decided to
treat them as true optional items. This difference between JCMT and
NRAO may have been driven by file format design given the difference
in approach between the self-describing GSD and the more statically
defined PDFL.  In retrospect it would have been better to attempt to
standardize even at the expense of having to spend more time in
discussion.

\subsection{Clear separation of model from file format}

GSDD benefited by explicitly defining the
data model for single-dish observing distinct from bytes on the
disk. However, whether by accident or design, the GSDD standard resulted in multiple
software implementations writing the data to disk in different formats
and using different techniques. The JCMT GSD format was never written
on anything other than a VAX but the NRAO format migrated from PDFL to
SDD going from VAX to Unix. Unfortunately these multiple formats also
meant that data reduction software
wishing to read the data would need to implement multiple file
readers. The reality is this work was never done. Given the focus of
both institutions on the use of GSDD in data acquisition using
different hardware platforms and different performance constraints,
this split is not surprising, but it is interesting to contemplate how
interoperability would have improved if the standards effort had also
included the definition of an interchange format. Being easily able to
compare a JCMT spectrum with a NRAO 12-m spectrum from within the same
data analysis package would have been extremely useful to the young
sub-mm community.

\subsection{A success apart}

Despite the lack of communication between implementors and the drift
in specifications, the GSDD format itself can be thought of as a
success when the uses of the format are looked at independently. The JCMT GSD
format was used for many years and files in this format are still
available. The related format continues to be used at the 12-m Telescope.

\subsection{Feeder for SDFITS}
\label{sec:sdfits}

GSDD was a very early attempt for independently funded and operated observatories to
agree on a shared data model. The goals of true interoperability of
raw telescope data amongst multiple data reduction software packages
was an important goal that was ahead of its time. Arguably the key
outcome of GSDD was that it motivated people to work together towards
a shared data format based on FITS. The GSDD experience fed in to a
1989 workshop held at Green Bank in late
1989\footnote{\url{http://fits.gsfc.nasa.gov/dishfits/dishfits.8910}}
that discussed how the community could migrate to a
single-dish FITS format. This was a key motivator for the adoption
of binary tables into the FITS standard \citep{1995A&AS..113..159C}
and ultimately led to the SDFITS standard \citep{2000ASPC..216..243G}.

\subsection{Communication}

A failing of GSDD is that when developers had real, practical reasons
to break a rule (e.g., needing a double precision word for a
pre-defined keyword when the standard required single precision, a
string needing 32 char instead of 16, changing the byte representation
from that of a VAX to IEEE), a forum had not been set up that could
negotiate modifications to the standard.  This is unlike the FITS
world where revisions to the definition have to pass through a
standards group.  A key lesson is that when a standard is
set up, the agreement should go beyond the expectation that ad hoc
conversations between staff at different observatories are a
sufficient means of keeping the standard viable.

The JCMT GSD library was documented and stable and the UK had the
Starlink Project \citep{1982QJRAS..23..485D} to publish the software
and data files to the UK community. However, access to that network from other
countries, such as the US, was problematic, and hindered the spread of
the software and prevented take up. Fears of lack of support also
drove people to create their own in-house solutions.

Today, 30 years on, the Internet and the culture of open-source
development make that much less likely and
and good ideas have a tendency to become distributed and generate a
supporting community outside of the original developers that ensures
its survival and growth.

\section{Thoughts on the Future}

Many of the lessons exposed by the history of GSDD have already been
learned in the 30 years since the key decisions were made and much
improved communications infrastructure has changed the way that people
work.  The current debate on future developments of data formats for
astronomy \citep[see e.g.][]{2015Thomas,2015Mink,2015MinkADASS}
indicates that there is a desire within the community for a format
that builds on the lessons learned using the FITS format to develop a
format with more modern underpinnings.  As noted in the debate
described in \citet{2015MinkADASS}, representing data on disk is
becoming a secondary concern relative to the discussion of data
models. A data model can be serialized into many different transport
and archive formats, and
it is relatively easy to make applications flexible enough to be
able to cope with these differences. Instead, it is much harder to
deal with different data models and implementation efforts should
concentrate on optimizing and generalizing the data model that is
being used. This is, after all,
the underlying \emph{business logic} that enables science to progress.
It may be true that all data models can be represented in a FITS file
but that doesn't mean that a FITS file is the most compact or most
efficiently accessed format. Changing the underlying file format used
in astronomy may simplify infrastructure libraries and result in new
abilities not available from within FITS. The easiest way to
migrate people to a new format may well be to do it without people
knowing what underlying format really is being used by their
applications.  As we move forward with discussions on data formats and
look again at hierarchical approaches
\citep[e.g.][]{2015Price,2015HDS,2015ASDF}, these may adjust the way
that people view data models. A hierarchical view is very different to
a flat view and data modelers should not be constrained by how their
models are represented on disk.

GSDD failed to unify the single-dish radio telescope community to use
a single file format. Focusing on the data model as a first step was
the correct decision at the time but it was poorly
implemented with little buy-in from the people writing the
software. Failing to agree on units, coordinate codes and the approach
to adding additional keywords removed any chance of GSDD being a
generically useful data model for the community.
Ideally a GSDD data model library should have been written to
abstract the file format completely from the user, but this was
all occuring before object-oriented programming was a common
paradigm. If GSDD were being implemented now it would be obvious how
to wrap data representing millimetre observations within object-oriented classes
involving differing receiver types and observing modes.

Abstracting the data model from the underlying file format is an idea
whose time has come. The Large Synoptic Survey Telescope data
management system \citep{2008arXiv0805.2366I,2010SPIE.7740E..1NK} uses
a \emph{butler} to mediate file access. The user requests data from
the system and the butler then pulls all the relevant data items
together (from a database or from files or from a combination of the
two) and instantiates an object representing that data. For LSST this
\texttt{Exposure} class represents something relevant to an optical
imager, but it could just as easily return an object that is relevant to
millimeter observing.

\section{Conclusions}

The GSDD data model was used at NRAO and JCMT for many years
but failed in its original goal of unifying single dish millimeter
astronomy and simplifying data reduction software reuse. As data
reduction packages have evolved it has become clear that the most
important aspect of such packages is format conversion such that the
software can map the external data model to an internal data model.
It is very hard to motivate individual observatories to target a
global standard for raw data without significant commitment and
obvious return on investement. NRAO and JCMT made a solid attempt but
could not maintain the momentum as other priorities intervened and
staff involved in the effort moved to other projects. Recent examples where observatories have
collaborated on a shared raw data format
\citep[e.g. MBFITS;][]{2006A&A...454L..25M} has shown that this is
possible but depends critically on the motivation of individuals and
on available funding\footnote{Ironically, the more money available the
  less chance there is an observatory will adopt a pre-existing system
  that might constrain their design decisions.}
 Interoperability of reduced data
products has significantly improved since the mid-1980s such that
there is a general expectation that reduced data cubes will be
viewable in general tools. By contrast, interoperability
of raw data has remained a much more elusive goal, at least amongst the
sub-mm radio telescope community.

\section*{Acknowledgments}

The National Radio Astronomy Observatory is a facility of the National
Science Foundation operated under cooperative agreement by Associated
Universities, Inc.
The James Clerk Maxwell Telescope has historically been operated by
the Joint Astronomy Centre on behalf of the Science and Technology
Facilities Council of the United Kingdom, the National Research
Council of Canada and the Netherlands Organisation for Scientific
Research. The JCMT part of this work was funded by the Science and
Engineering Research Council and subsequently the Particle Physics and
Astronomy Research Council.
We thank Thomas Folkers and Harvey Liszt for useful
discussions on GSDD. This research has made use of NASA's Astrophysics
Data System.

The source code for the JCMT GSD library (C, Perl wrapper, and Java)
and the source code for UniPOPS are available from
Github\footnote{UniPOPS at \url{https://github.com/nrao/UniPOPS} and JCMT
  GSD at \url{https://github.com/Starlink/starlink}}
and both are distributed under the Gnu General Public Licence. The
JCMT GSD library is distributed as part of the Starlink software
collection \citep[see e.g.,][\ascl{1110.012}]{2014ASPC..485..391C}.
Some of the harder to find documents referenced by this paper have
been collected and made available on Github at
\url{https://github.com/timj/aandc-gsdd/tree/master/support_docs}.

\end{document}